\documentclass[12pt]{article}
\usepackage[latin1]{inputenc}
\usepackage[english]{babel}
\usepackage{graphicx}
\usepackage{color}
\usepackage{epsfig}
\usepackage{amssymb}
\usepackage{amsmath}
\usepackage{a4wide}
\usepackage{verbatim}
\usepackage{fancyhdr}
\usepackage{subfigure}
\usepackage{float}
\usepackage{fixltx2e}
\usepackage[ps2pdf]{thumbpdf}
\usepackage[ps2pdf]{hyperref}
\usepackage[numbers]{natbib}
\usepackage{setspace}
\usepackage{hhline}
\usepackage{authblk}

\begin{document}

\newcommand{\Angst}{$\mathring{\mathrm{A}}$}
\newcommand{\DEG}{$^\circ$}
\newcommand{\sq}{$^\mathrm{2}$} 
\newcommand{\NIMA}{Nucl. Instrum. Meth. A}
\newcommand{\extB}{\textit{extended kinked ballistic double-ellipse B} }

\thispagestyle{plain}
\date{}
\title{An improved elliptic guide concept for a homogeneous neutron beam without direct line of sight}
\author[1,2,*]{C. Zendler}
\author[1,2]{D. Nekrassov}
\author[1,2]{K. Lieutenant}
\affil[1]{Helmholtz-Zentrum Berlin f{\"u}r Materialien und Energie GmbH, Hahn-Meitner-Platz 1, D-14109 Berlin, Germany}
\affil[2]{ESS Design Update Programme, Germany}
\affil[*]{\textbf{corresponding author}: carolin.zendler@esss.se}
\maketitle

\begin{abstract}
Ballistic neutron guides are efficient for neutron transport over long distances, and in particular elliptically shaped guides have received much attention lately. However, elliptic neutron guides generally deliver an inhomogeneous divergence distribution when used with a small source, and do not allow kinks or curvature to avoid a direct view from source to sample. In this article, a kinked double-elliptic solution is found for neutron transport to a small sample from a small (virtual) source, as given e.g. for instruments using a pinhole beam extraction with a focusing feeder. A guide consisting of two elliptical parts connected by a linear kinked section is shown by VITESS simulations to deliver a high brilliance transfer as well as a homogeneous divergence distribution while avoiding direct line of sight to the source. It performs better than a recently proposed ellipse-parabola hybrid when used in a ballistic context with a kinked or curved central part. Another recently proposed solution, an analytically determined non-linear focusing guide shape, is applied here for the first time in a kinked and curved ballistic context. The latter is shown to yield comparable results for long wavelength neutrons as the guide design found here, with a larger inhomogeneity in the divergence but higher transmission of thermal neutrons. It needs however a larger (virtual) source and might be more difficult to build in a real instrument.

\end{abstract}



\section{Introduction}

Neutron guides are important tools used to deliver ample flux to samples at large distances from the source. Long neutron beamlines lead to low background, and are of particular interest to the planned European Spallation Source (ESS)~\cite{ESS_TDR} due to its long pulse and required time of flight resolution. Ballistic guides in which an expanding guide section reduces the beam divergence before the neutrons are transported by a straight guide and then focused by a guide section of decreasing spatial extension have been shown to perform better than conventional straight or curved neutron guides \cite{MezeiBallistic}, and elliptic or parabolic guide shapes can improve the transmission even further \cite{KlenoBallisticGuides}.

Hence a currently widely studied guide shape is the elliptic guide profile, which in principle allows neutrons from one focal point to be transmitted to the second focal point with just one reflection. This idealized behavior was recently shown to be true only for a negligibly small fraction of neutrons under realistic conditions \cite{MultipleReflectionsInEllipticNeutronGuides}. Even though the beam homogeneity after elliptically focusing is in general superior to the one after linear or parabolically focusing guides \cite{Schanzer200463}, inhomogeneous divergence distributions are often seen in Monte Carlo simulation studies of elliptical guides \cite{MultipleReflectionsInEllipticNeutronGuides,StahnSelene1,BentleyHybrid}. A further drawback is the difficulty to avoid direct line of sight, which can be solved in certain cases by gravitational bending of the elliptical guide \cite{GravitationalCurving}, but this solution is limited to long wavelengths in a small waveband around the wavelength for which it was optimized, and cannot be expected to reduce any inhomogeneities of the divergence spectrum. An alternative central beamstop to block the direct line of sight will create a hole in the transmitted phase space.

A double ellipse somewhat improves the divergence profile with only a small loss in transmission \cite{MultipleReflectionsInEllipticNeutronGuides} if the two ellipses have the same characteristic angle $\psi=\arctan{\left(b/a\right)}$, with $a$ and $b$ the long and short half axis length, and share a common central focal point. Furthermore, a double elliptic guide design provides a natural narrow point, which constitutes a second eye of the needle for a possible chopper placement \cite{Boeni20081}. This principle is used in the Selene~\cite{StahnSelene} and POWTEX~\cite{Powtex} guide concepts, achieving a homogeneous divergence in the respective simulations. While the latter does not avoid direct line of sight, the former does so by using only quarter ellipses acting as elliptical mirrors rather than guides, which are inclined and combined with slits and shielding equivalent to an effective central beamstop. This approach sacrifices intensity on the sample because the design is optimized for low background, using only the small fraction of ideal neutron trajectories from an approximate point source. Its use of elliptical guide parts as focusing devices further constrains it to shorter instrument in which gravity effects are small.

This article describes a kinked double-elliptic guide concept for a 150\,m long instrument looking at a small source, which focuses the neutron beam onto a small sample without introduction of beam inhomogeneities. The source can either be a real moderator, or a virtual source created by a preceding slit or focusing device as often considered for beamlines at the ESS, where a small beam spot close to the source is advantageous for the use of a pulse shaping chopper as well as for the placement of shielding. Guide systems intended for direct transportation from large sources will be discussed elsewhere. 

Based on the ideas of L. D. Cussen \cite{EG15,TalkLeosGuideIdea,Talk2Essex}, the motivation for a kinked ballistic double-ellipse, consisting of two elliptically shaped guide parts in the beginning and end connected by a linear guide section, is recalled in section~\ref{sec_Idea}. After a description of boundary conditions and simulation details in section~\ref{sec_SimDetails}, the principles of the argumentation are verified by simulation and used to design a modified kinked ballistic double-ellipse in section~\ref{sec_Simulation}. This new guide design is then compared in section~\ref{sec_AlternativeIdeas} to two recently proposed alternative approaches that have been shown to give a homogeneous divergence distribution in a focused neutron beam under different conditions or in theory: First, for the special case of a large source and an ellipse with a large opening, a hybrid guide consisting of an elliptically diverging and a parabolically converging part with roughly equal lengths and a total length of about 50\,m has been shown to yield an improved divergence profile compared to a full ellipse \cite{BentleyHybrid}. Second, an analytical calculation using phase space considerations resulted in a non-linear shape for a focusing guide that retains a rectangular phase space \cite{NonLinearFocusingGuide}. These approaches are investigated here under the condition of a small (virtual) source, characterizing their performance in a ballistic guide design including a kinked or curved section to avoid direct line of sight.

\section{Theoretical considerations}\label{sec_Idea}
This section derives a double-ellipse with a linear kinked connecting section and central narrow point from ellipse properties in combination with non-ideal behavior of neutrons not coming from a focal point, summarizing considerations made in \cite{EG15,TalkLeosGuideIdea,Talk2Essex} in context of a guide design for an ESS extreme environment (ESSEX) instrument.

\subsection{General guide shape}

\begin{figure}[tb]
\centering
\subfigure[\label{f_SkizzeG2}]{\includegraphics[width=.3\linewidth]{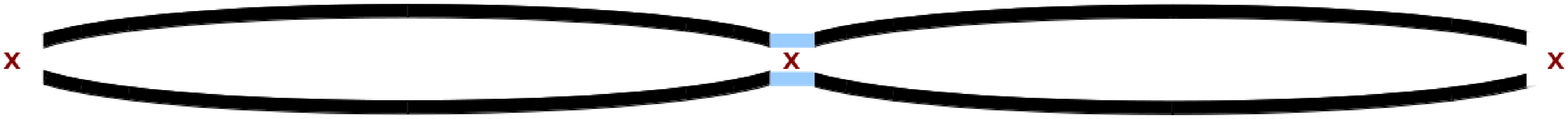}}
\subfigure[\label{f_SkizzeG3}]{\includegraphics[width=.3\linewidth]{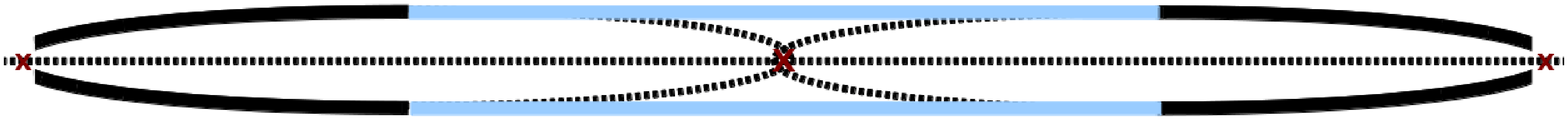}}
\subfigure[\label{f_SkizzeG6}]{\includegraphics[width=.3\linewidth]{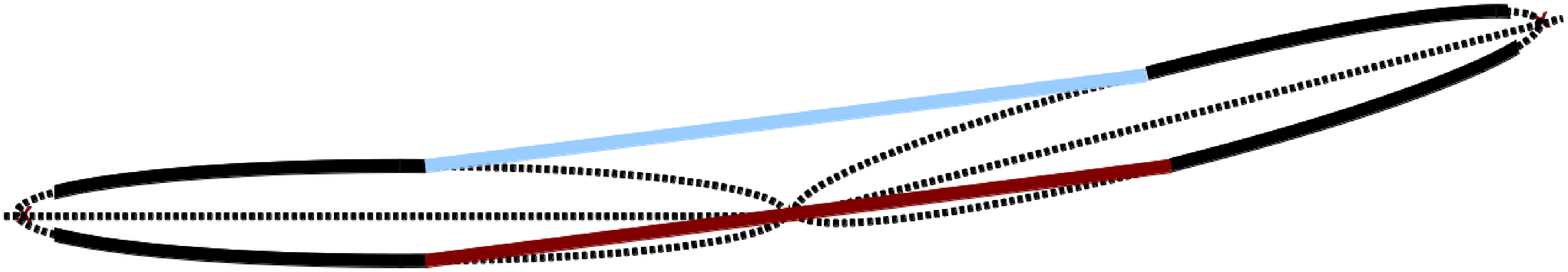}}
\subfigure[ \label{f_SkizzeG7}]{\includegraphics[width=.3\linewidth]{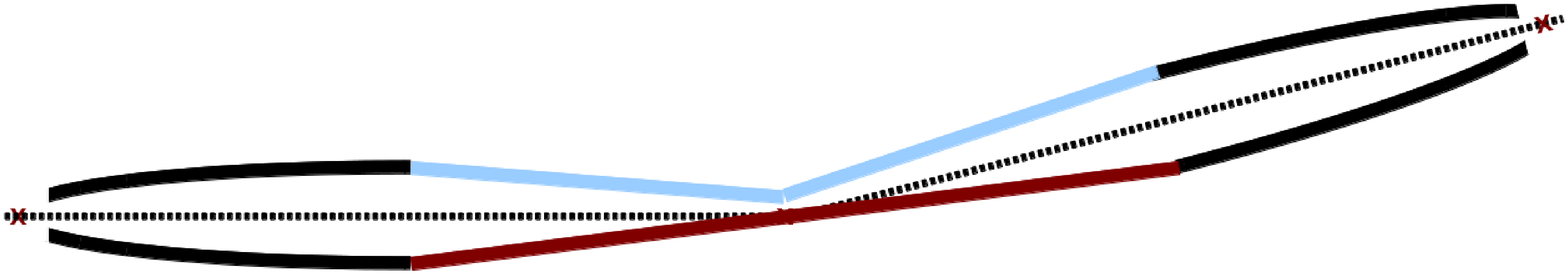}}
\subfigure[ \label{f_SkizzeF9}]{\includegraphics[width=.3\linewidth]{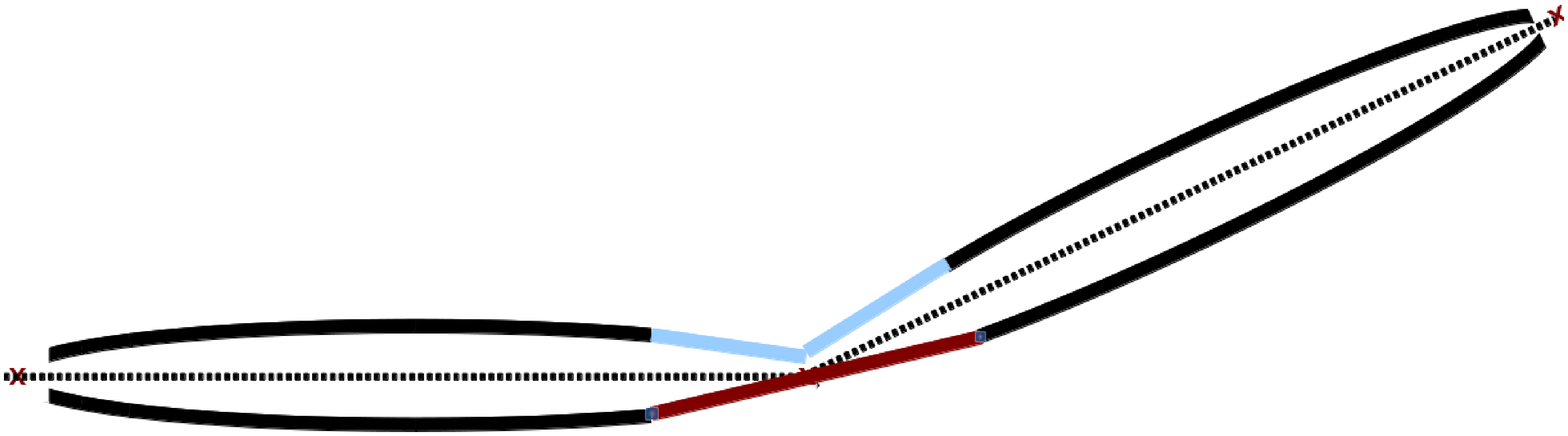}}
\subfigure[\label{f_SkizzeFinal}]{\includegraphics[width=.3\linewidth]{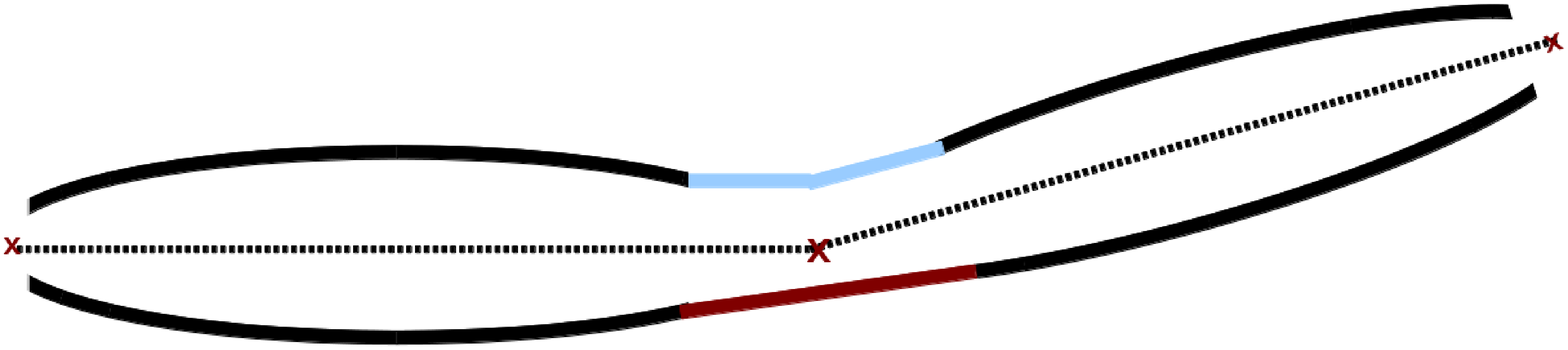}}
\caption{Derivation of a kinked ballistic double-ellipse with narrow point, schematic drawing. (a) double-ellipse (b) ballistic double-ellipse (c) simple kinked ballistic double-ellipse (d) kinked ballistic double-ellipse with narrow point (e) extended kinked ballistic double-ellipse (f) extended kinked ballistic double-ellipse B }
\label{f_Skizzen}
\end{figure}

The second ellipse in a double-elliptic guide as drawn in figure~\ref{f_SkizzeG2} principally reverses neutron trajectories and thereby partly reverses unwanted aberration effects of the first ellipse only under idealized conditions of neutrons emerging from a point source. Under realistic conditions, i.e. with an extended source (millimeters and larger), gravity altering neutron trajectories (and thereby reflection angles) and a possible non-perfect elliptic guide shape constructed out of straight pieces, most neutrons will undergo multiple reflections already in the first ellipse and the neutron beam cannot be expected to be perfectly focused at the central focal point. The transmission compared to a single ellipse will decrease even more than expected from the ideally only doubled number of reflections. Hence a straight forward modification to the double elliptic guide design is to replace the converging part of the first ellipse as well as the diverging part of the second ellipse by a straight guide, to obtain the ballistic double-ellipse in figure~\ref{f_SkizzeG3}. Note that differently to most ballistic guides with elliptically focusing ends, the elliptic guide parts still share a common focal point in the center of the straight guide. 

\subsection{Avoiding direct line of sight}

The constant connection between elliptic guide parts in the ballistic double-ellipse allows to introduce two kinks at the connection points. Most neutrons enter the guide with a divergence large enough to undergo at least one reflection in the first half of the ellipse, leading to a smaller divergence in the central part of the guide and making this an advantageous position to place a mirror to reflect the neutron beam out of direct line of sight. The kink angles are demanded in \cite{EG15,TalkLeosGuideIdea,Talk2Essex} to be such that the common central focal point lies on the reflecting guide wall, as illustrated in figure~\ref{f_SkizzeG6}, and a central narrow point as shown in figure~\ref{f_SkizzeG7} is introduced to provide a suitable position for frame overlap choppers and to be further out of line of sight. The study in \cite{EG15,TalkLeosGuideIdea,Talk2Essex} further states that extending the used part of the ellipses while narrowing the maximal guide width improves the divergence profile further. An example of such a guide is schematically shown in figure~\ref{f_SkizzeF9} and will be referred to as \textit{extended kinked ballistic double-ellipse A}.

This design is improved here to an \textit{extended kinked ballistic double-ellipse B} schematically drawn in \ref{f_SkizzeFinal}: The condition of the kinked guide wall going through the central focal point causes kink angles larger than necessary to avoid direct line of sight, therefore this condition is dropped. The guide width at the central narrow point, which is fixed in \cite{EG15,TalkLeosGuideIdea,Talk2Essex} to 4\,cm, will here be chosen such that the guide walls opposite to the mirror wall are parallel to the guide axes. This is illustrated by the light blue lines in figure \ref{f_SkizzeFinal}. This way the central guide width as well as the kink angle are determined by the transition point between ellipse and linear guide. The optimal transition point is found by simulation in section~\ref{sec_Simulation}.

\section{Boundary conditions and simulation details}\label{sec_SimDetails}

The distance from the source to the 1$\times$1\,cm\sq\,sample is fixed to 150\,m. A dedicated study of pinhole beam extraction \cite{PinholeBeamExtraction} revealed that efficient neutron transport as well as a satisfying beam homogeneity with an ellipse in a 150\,m long instrument can be achieved using a 3$\times$3\,cm\sq\,pinhole for a desired divergence of $\pm$1\DEG, and a smaller (larger) pinhole size for smaller (larger) divergence. In order to cover a somewhat larger divergence here, the source size is initially chosen to be 4$\times$4\,cm\sq. The effect of a smaller source is shown in section~\ref{s_SourceSize}. For general results independent of a certain source spectrum shape, the simulated source emits a normalized rectangular wavelength spectrum of 1\,cm$^{\mathrm{-2}}$s$^{\mathrm{-1}}$str$^{\mathrm{-1}}$\Angst$^{\mathrm{-1}}$.   

Guide kinks or curvature are introduced only in the horizontal dimension so their effects can be studied independently of gravity. The guides have rectangular cross-sections, and the entry and exit width $w_{in,out}$ is 2\,cm and hence sufficiently small compared to the source size, which has been shown to be a precondition for avoiding irregularities in the divergence caused by the elliptical guide shape itself \cite{MultipleReflectionsInEllipticNeutronGuides} or gravity \cite{GravityInEllipticGuides}. The distance from source to guide entry position $d_s$ is the same as the distance from guide exit to sample position. 

The central narrow point is initially chosen to be of the same size as the (virtual) source. 

The Monte Carlo ray tracing simulation package VITESS~\cite{Vitess} version 3.0 was used to simulate the data. Elliptic guide profiles are modeled as perfect ellipses without taking guide segmentation into account because segmentation or misalignment effects are not the topic of this work. To minimize losses in the thermal neutron range and otherwise keep the study general, all simulations use $m=$\,6 mirror coatings, where the $m$ number refers to the supermirror coating with a critical angle for total reflection of $m$ times the critical angle of nickel. In a practical guide design, the coating has to be optimized for the desired wavelength and divergence on the sample, and a much lower coating can be expected to be sufficient in most regions of the guide \cite{Powtex}.

\section{Simulation results} \label{sec_Simulation}

\begin{figure}[tb!]
\centering
\subfigure[\label{f_Parameter}]{\includegraphics[width=.49\linewidth]{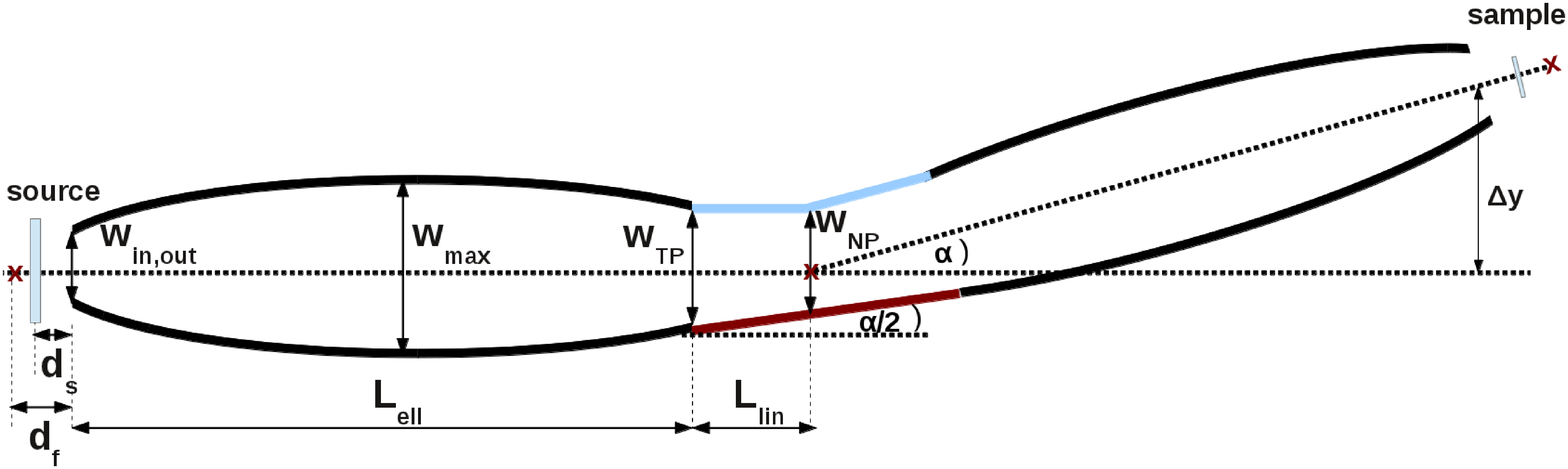}}
\subfigure[\label{t_Parameter}]{
\begin{footnotesize}
\begin{tabular}{||c|c|c|c|c|c|c|c|c|c|c||}
\hhline{===========}
 guide &  $w_{max}$  &  $w_{TP}$ & $w_{NP}$  & $w_{in,out}$   & $d_s$   & $d_f$   & $L_{ell}$   &  $L_{lin}$  & $\alpha$ & $\Delta y$   \\
       &       [cm]  &        [cm] &    [cm] &         [cm]  &   [cm]  &    [cm] & [m]        &      [m]              &  & [cm]  \\ 
\hline
ext. knkd dbl-ell (A) & 10 & 8  &  4.0 & 2 &  19  & 76   & 59.8 & 15.0 & 0.30\DEG & 40 \\ 
ext. knkd dbl-ell (B) & 15 & 8  &  7.5 & 2 &  19  & 34   & 69.0 & 5.8  & 0.11\DEG & 14 \\
\hhline{===========}
 guide &  $w_{max}$  &  $w_{TP}$ & $w_{eff}$  & $w_{in,out}$   & $d_s$   & $d_f$   & $L_{b}$   &  $L_{c}$  & $R$ & $\Delta y$   \\
       &      [cm]  &      [cm] &    [cm]   &         [cm]  &   [cm]  &    [cm] & [m]        &      [m]              & [m] & [cm]  \\
\hline
curved hybrid       & 15 & 15   & 10.5      & 2             &  19     &  34     & 37.6       & 74.5    & 19946\,m  & 28 \\
curved non-linear   & 10 & 10   &  8.0      & 2             &  19     &  -      & 19.3       & 111.1   & 32687\,m  & 25 \\
\hhline{===========}
\end{tabular}
\end{footnotesize}
}
\caption{Schematic drawing (a) and values (b) of guide parameters in horizontal plane: maximal guide width $w_{max}$, width at transition point $w_{TP}$ and narrow point $w_{NP}$, entry and exit width $w_{in,out}$, distance from guide to source and sample $d_s$ and to focal points $d_f$, length of one elliptic part $L_{ell}$ and linear or constant length between ellipse and central focal point $L_{lin}$, total kink angle $\alpha$ and total horizontal shift $\Delta y$. In the curved hybrid and non-linear designs shown for comparison, the curvature radius $R$ and the effective guide width $w_{eff}$ are given instead of the kink angle and guide width at narrow point, $L_c$ is the length of the curved central section and $L_b$ the length of one ballistic guide section (elliptic, parabolic or analytic non-linear shape).}
\label{t_vglG5G7}
\end{figure}

In this section, the considerations made in section~\ref{sec_Idea} are first verified by simulation, before an improved guide called the \extB is designed. Figure~\ref{f_Skizzen} shows the gradual construction of this guide starting from a double-ellipse. Its parameters are given in the second row of table~\ref{t_Parameter} (\textit{ext. knkd. dbl-ell (B)}). The same ellipse parameters have been used in all the simulations of the double-elliptic guides shown in figure~\ref{f_Skizzen}, other parameters are chosen as described in section~\ref{sec_SimDetails}. A single ellipse with two times the major and minor semi-axes has been simulated for comparison. 

The divergence profile is smoothened when moving from a single elliptic guide to a double ellipse (fig.~\ref{f_SkizzeG2}) and further to the ballistic double-ellipse design (fig.~\ref{f_SkizzeG3}). However, a substantial loss of intensity on the sample (with the settings used here, 35\%-50\% loss depending on the neutron wavelength) is seen with the double-elliptic design compared to a single ellipse which is only partly re-gained in the ballistic guide, in which especially short wavelengths benefit from the removal of the narrow central part. 

\begin{figure}[tbh!]
\centering
\subfigure[\label{f_HorDivG6}]{\includegraphics[width=.45\linewidth]{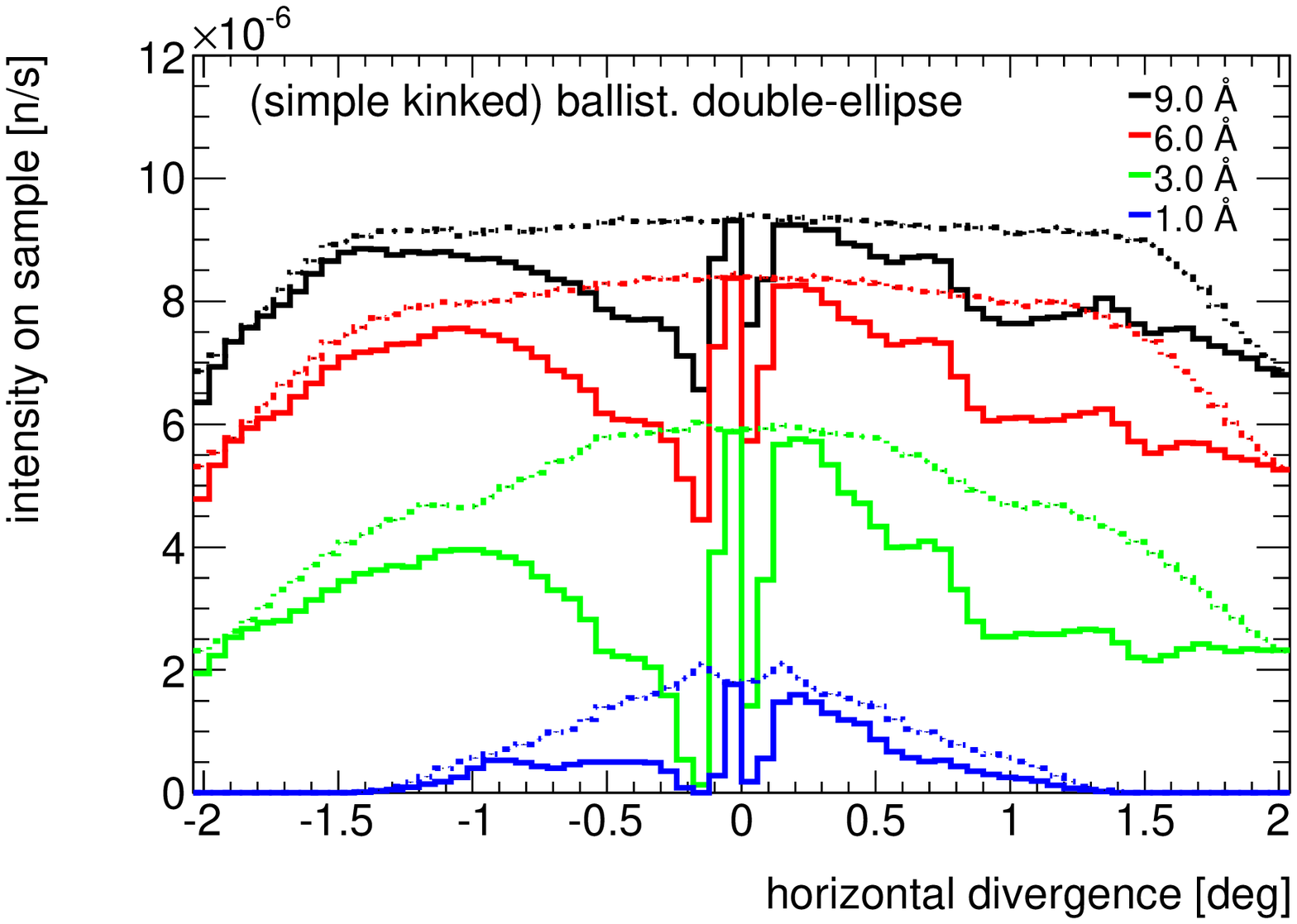}}
\subfigure[\label{f_BTG3}]{\includegraphics[width=.45\linewidth]{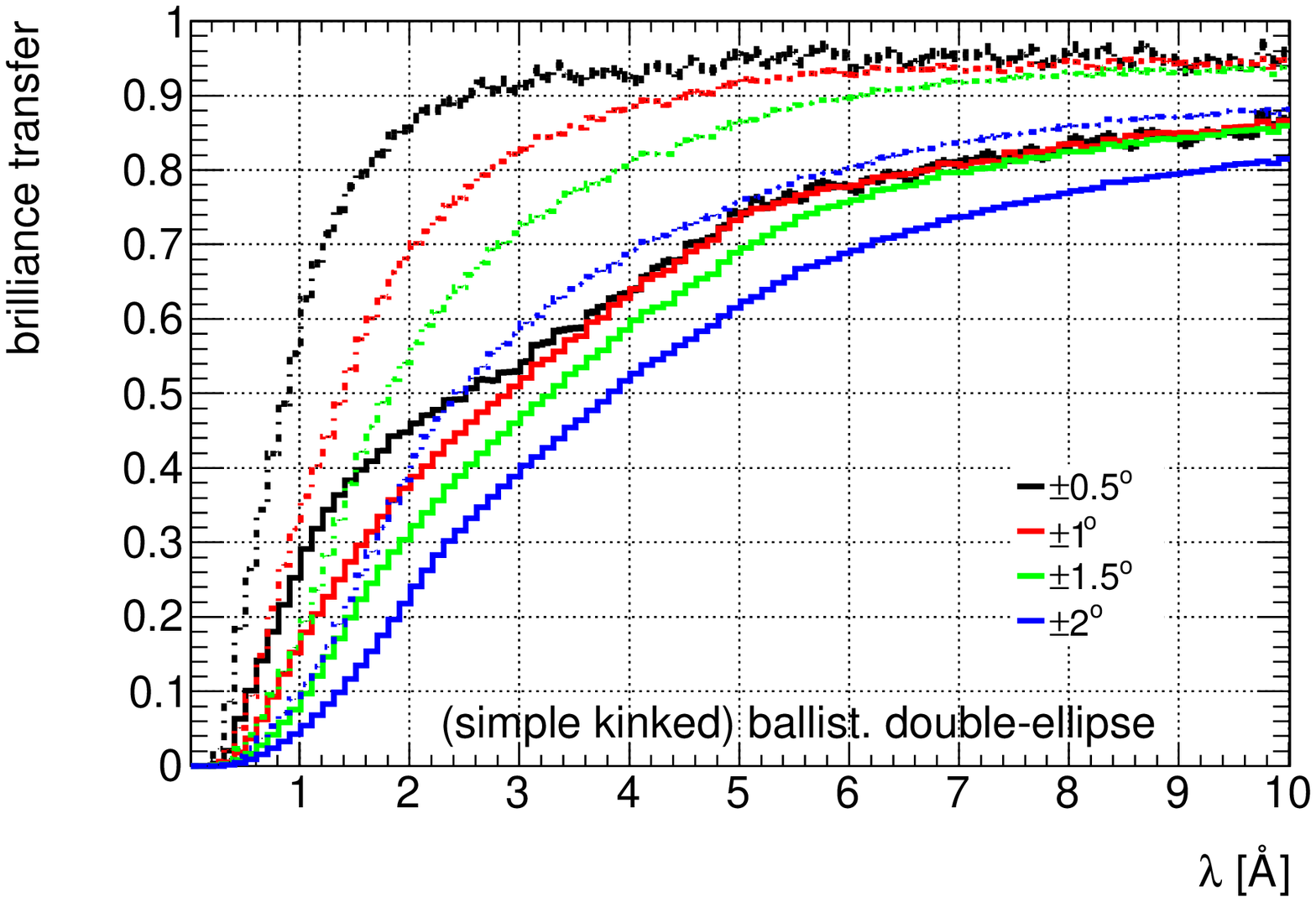}}
\caption{Horizontal divergence in bins of 0.06\DEG\,for different wavelengths (a) and brilliance transfer for different divergence regions (b), obtained with the \textit{(simple kinked) ballistic double-ellipse} in dotted (solid) lines. Error bars are drawn but mostly too small to see, and are calculated from the generated statistics of neutron trajectories.}
\label{f_vglG5}
\end{figure}

Kinking the ballistic double-ellipse as in figure~\ref{f_SkizzeG6} causes deep breaches in the divergence profile, which are shown as solid lines in figure~\ref{f_HorDivG6} for the four different wavelengths 1\,\Angst, 3\,\Angst, 6\,\Angst\,and 9\,\Angst\,in bands of $\pm$0.05\,\Angst. Only neutrons on the 1$\times$1\,cm\sq\,sample are shown. The performance of the un-kinked version of this guide (as in figure~\ref{f_SkizzeG3}) is added as dotted lines for comparison, showing hardly any structure in the divergence distributions. The brilliance transfer (BT) is shown in figure~\ref{f_BTG3}, where brilliance transfer is defined as the ratio of neutron flux per solid angle and wavelength on the sample compared to the source position. Four different divergence ranges are treated, where neutrons outside the given divergence limits are ignored in the calculation of BT, i.e. suppression of unwanted neutrons with too large divergences is not included here. With this definition, a BT of 80\,\% is reached for neutrons with divergence $<$\,0.5\DEG (1\DEG, 1.5\DEG, 2\DEG) and $\lambda >$\,1.6\,\Angst\,(2.7\,\Angst, 3.9\,\Angst, 5.9\,\Angst) in the un-kinked version of the ballistic guide (dotted lines). Comparing solid and dotted lines in figure~\ref{f_BTG3} illustrates the transmission loss caused by the kink.

In the shown examples, transmission of thermal neutrons has been enhanced by moving the outer focal points slightly beyond the source and sample position, such that for a fair comparison, the ellipse parameters are identical to the ones in the \extB solution found later. The kink angle obtained when placing one guide wall in the common central focal point is not large enough to loose direct line of sight to the source without the central narrow point (fig.~\ref{f_SkizzeG6}), therefore a slightly larger angle has to be used. 

\begin{figure}[tbh!]
\centering
\subfigure[\label{f_HorDivG7}]{\includegraphics[width=.45\linewidth]{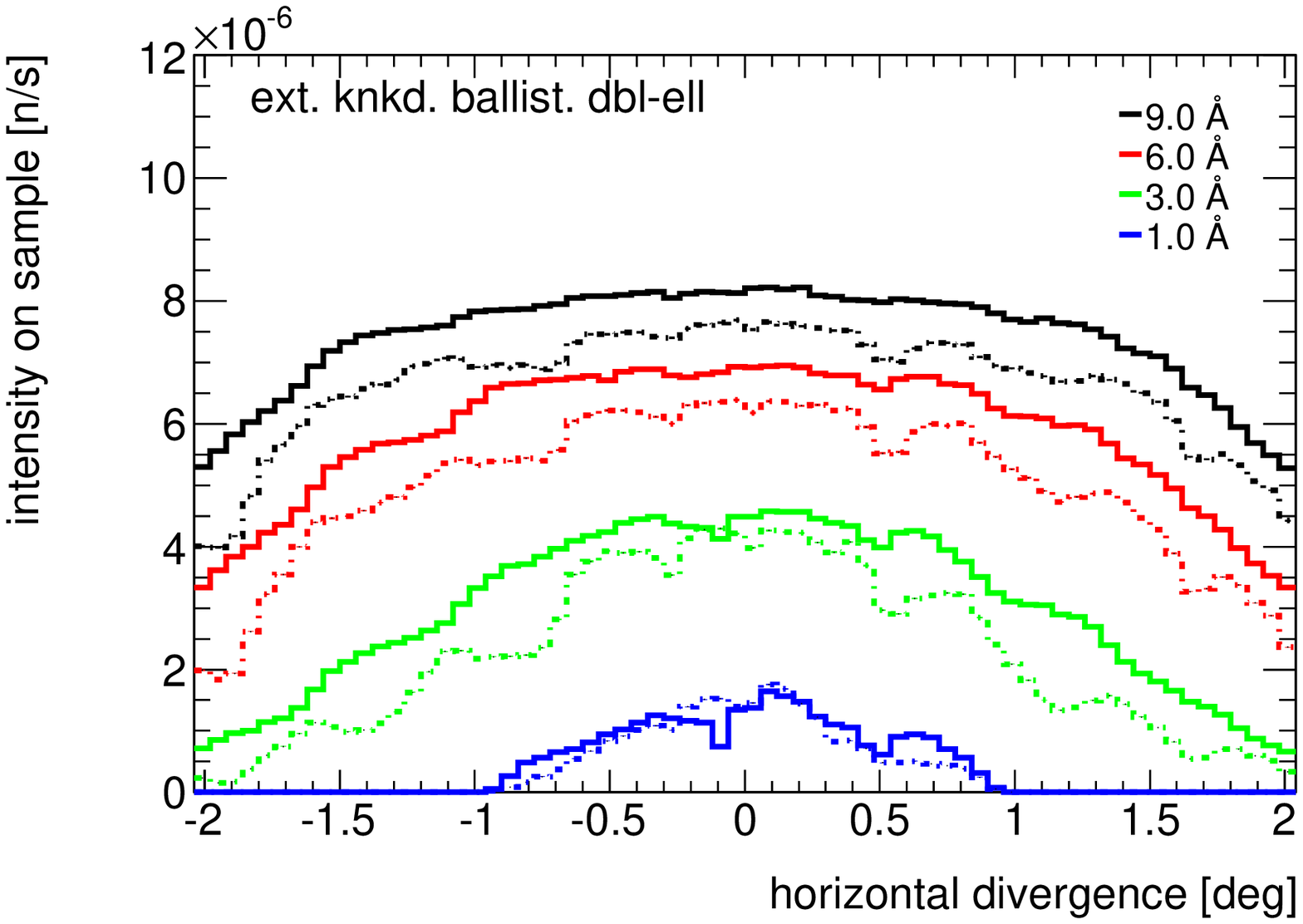}}
\subfigure[\label{f_BTG7}]{\includegraphics[width=.45\linewidth]{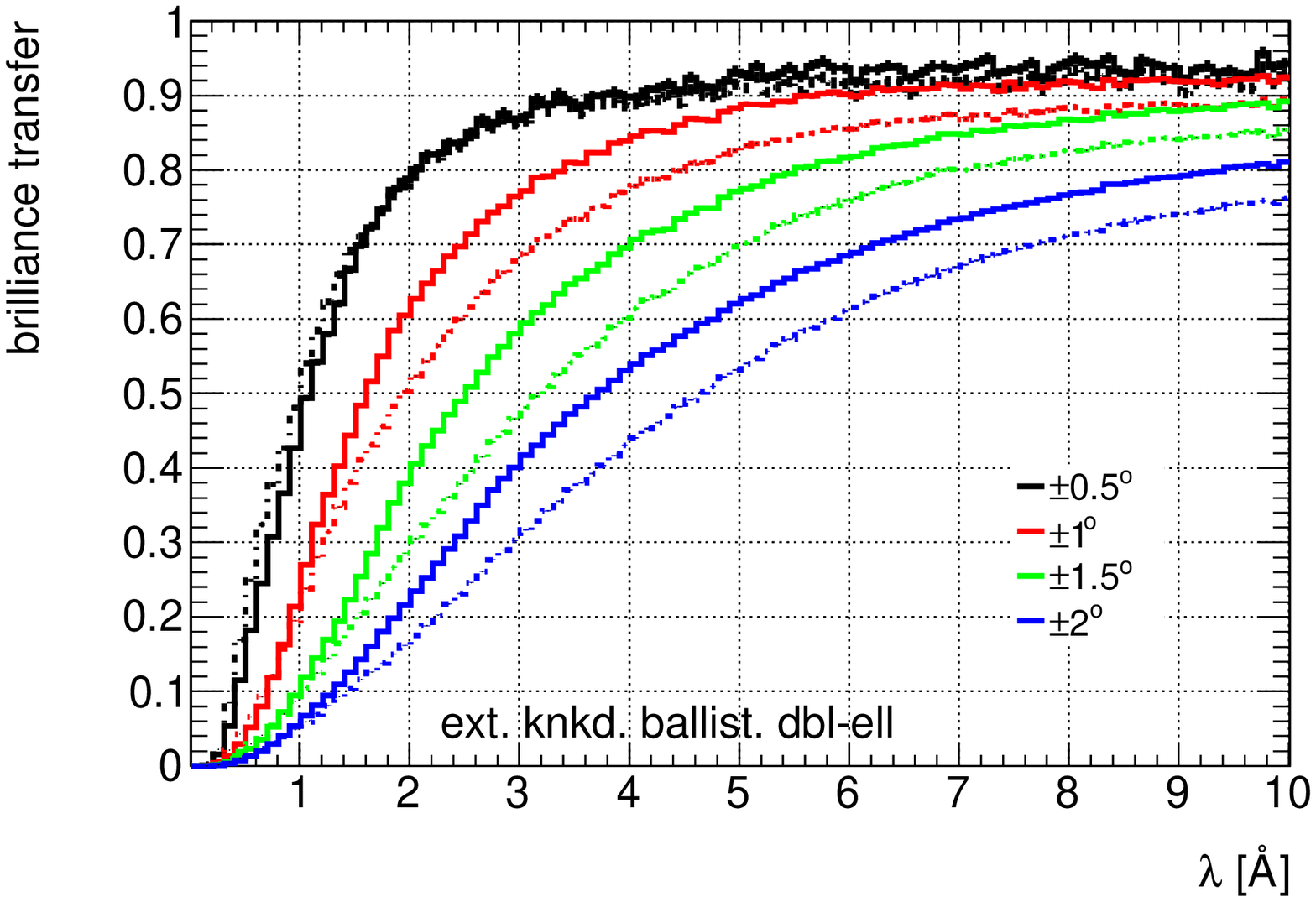}}
\caption{Horizontal divergence in bins of 0.06\DEG\,for different wavelengths (a) and brilliance transfer for different divergence regions (b), obtained with the \textit{extended kinked ballistic double-ellipse A} (dotted) and \textit{B} (solid). Error bars are drawn but mostly too small to see, and are calculated from the generated statistics of neutron trajectories.}
\label{f_vglG7}
\end{figure}

The introduction of a narrow point as in figure~\ref{f_SkizzeG7} reduces the needed kink angle and allows to place one guide wall on the central focal point. However, this smoothens the divergence profile only marginally, and leads to large losses in the short wavelength regime. By extending the elliptical shape beyond the position of maximal guide width as in figure~\ref{f_SkizzeF9}, the divergence distribution can be further smoothened but still shows some structure. Only if the maximal guide width is limited to 10\,cm, a greatly improved divergence profile is achieved, which is shown by the dotted lines in figure~\ref{f_HorDivG7}. As in \cite{EG15,TalkLeosGuideIdea,Talk2Essex}, the ellipse has been extended up to 80\,\% here, corresponding to a guide width of 8\,cm at the transition point from elliptic to linear shape. There is, however, still some structure visible, and the reduction of maximal guide width has to be paid by a decreased transmission. 

If the condition of one guide wall passing through the central narrow point is dropped, the kink angle can be reduced to a value needed to prevent direct line of sight. As a consequence, the divergence profile becomes smoother. The reduction of the maximal guide width can be undone without re-introducing inhomogeneities, thus enhancing transmission. The optimal transition point between elliptic and linear guide is found by a simulation scan to be still at 8\,cm guide width, which corresponds to building 92\,\% of the ellipse. The guide is furthermore wider at the narrow point due to the construction outlined in section~\ref{sec_Idea}. This modified design will be called \textit{extended kinked ballistic double-ellipse B}. Its performance is shown by the solid lines in figures~\ref{f_HorDivG7} and \ref{f_BTG7}: both transmission and divergence profile are improved compared to the former guide design \textit{A}. For 1\,\Angst\,neutrons or those with very small divergence, the result is about the same. Only for short wavelength neutrons with small divergence, design \textit{A} gives a slightly higher transmission, which is however caused by the smaller guide width and not by the design principle. With design \textit{B}, a BT of 80\,\% is reached for neutrons with divergence $<$\,0.5\DEG (1\DEG, 1.5\DEG, 2\DEG) and $\lambda >$\,2\,\Angst\,(3.5\,\Angst, 5.5\,\Angst, 9.5\,\Angst). 

The parameters of both \textit{extended kinked ballistic double-ellipse} versions \textit{A} and \textit{B} are given in the first two rows of table~\ref{t_Parameter}. In the vertical plane, the elliptical guide parts are the same as the horizontal design but instead of the linear kinked guides in between, a constant connection with height $h_{vert}=w_{TP}$ is used.

\section{Comparison with hybrid and analytic non-linear approach}\label{sec_AlternativeIdeas}

Two other solutions to the problem of an inhomogeneous divergence profile of a focused neutron beam have recently been proposed. These are adapted to the boundary conditions of a 4$\times$4\,cm\sq\,source, a 150\,m long instrument and the avoidance of direct line of sight in this section. Their performance in a ballistic guide design with parameters similar to the ones used in the \extB solution described above is then compared. Both a kinked and a curved central section are studied.

\subsection{Ellipse-parabola hybrid}

An elliptically diverging guide followed by a parabolically focusing guide was shown in \cite{BentleyHybrid} to yield a smoother divergence distribution than a comparable full ellipse. The cited study uses an 18.6\,cm wide source, a 4$\times$4\,cm\sq\,sample and a 50\,m long guide. In order to match the requirements used here, a guide similar to the \textit{kinked ballistic double-ellipse with narrow point} (cf. figure~\ref{f_SkizzeG7}) is designed which in principal is identical to the solution found above but with a transition point at the maximum of the ellipse. The focusing ellipse is replaced by a parabola with the same exit window of 2$\times$2\,cm\sq. This places the focal point of the parabola about 50\,cm behind the sample. The source size is again 4$\times$4\,cm\sq. 

\begin{figure}[tb!]
\centering
\subfigure[\label{f_HorDivHyb}]{\includegraphics[width=.45\linewidth]{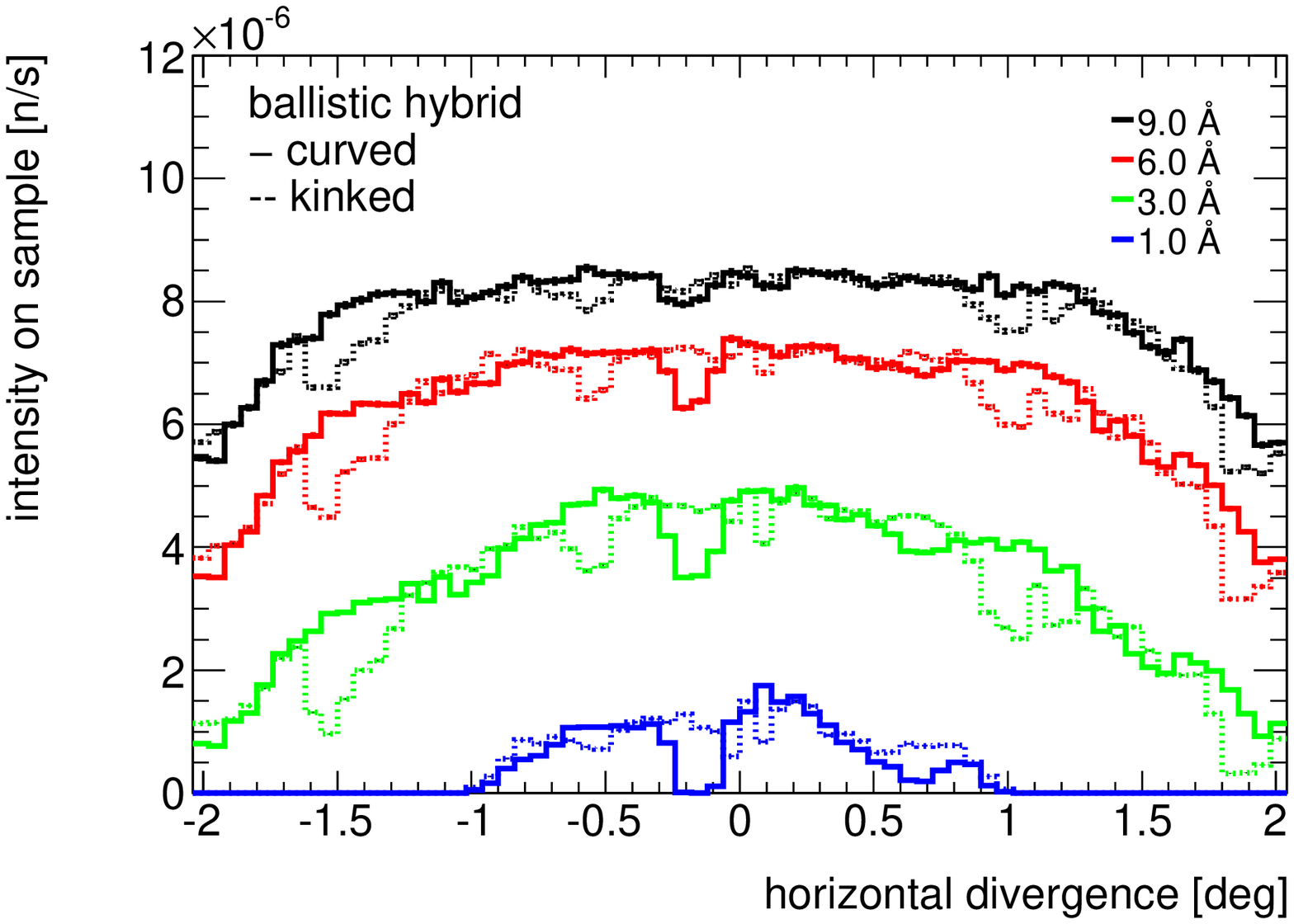}}
\subfigure[ \label{f_BTratio_Hybrid}]{\includegraphics[width=.42\linewidth]{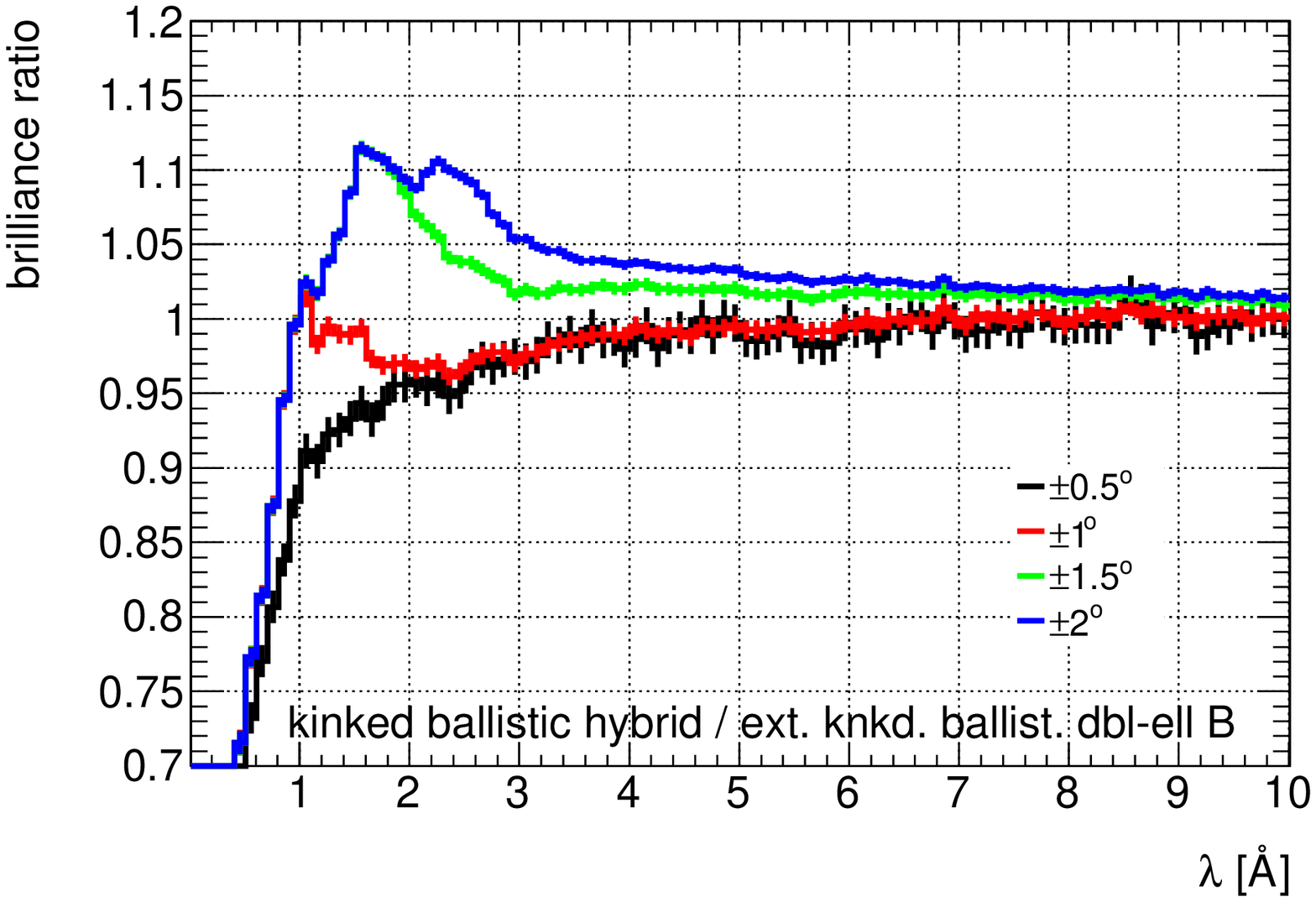}}
\caption{Horizontal divergence in bins of 0.06\DEG\,obtained with curved (solid) and kinked (dotted) hybrid guide design (a) and brilliance for different divergence regions divided by the one obtained with the \textit{extended kinked ballistic double-ellipse B} solution (b).}
\label{f_Hybrid}
\end{figure}

Both kinked and curved version of the ballistic hybrid guide show a structure in the divergence profile in figure~\ref{f_HorDivHyb}, with local minima at different divergence values and an overall irregular divergence distribution for short wavelengths. In terms of beam homogeneity, the hybrid type solutions cannot compete with the \textit{extended kinked ballistic double-ellipse B}. The transmission is however slightly higher for large divergence ($> \pm$1.5\DEG) neutrons and $\lambda >$\,1\,\Angst, as can be seen in figure~\ref{f_BTratio_Hybrid} showing the ratio of brilliance on the sample obtained with the kinked hybrid compared to the \extB solution found above. For small divergence up to 1\DEG, the transmission is slightly lower. Apart from very short wavelengths $<$1\,\Angst, the maximal difference in brilliance transfer is of the order of 10\,\%, so the overall transmission of the hybrid solutions is comparable to the one of the \textit{extended kinked ballistic double-ellipse B}, while the beam homogeneity is lower.

\subsection{Analytically calculated non-linear focusing}

An analytical calculation~\cite{NonLinearFocusingGuide} yielded a solution for a focusing guide that retains a rectangular phase space and hence a homogeneous divergence distribution. The proposed focusing guide consists of a non-linear section followed by a linear section with wall inclinations identical to the angle at the exit of the non-linear part. The shape of the non-linear guide walls depend on the entry width, the desired ratio of guide width change to entry width (compression), the target wavelength as well as the input divergence, the latter being equivalent to the m-number of a preceding straight guide. Keeping a desired exit width of 2\,cm, the maximal guide width is set to 10\,cm since irregularities occur for larger compression factors. The target wavelength is set to 0.9\,\Angst\,in the calculation to ensure a good transmission of 1\,\Angst\,neutrons, the coating in the straight section is assumed to be $m=$\,2. These settings lead to a necessary coating of $m=$\,5.2 in the non-linear guide section. These coating conditions are just used to calculate the shape of the guide; for the sake of comparability with the previous guide simulations, $m=$\,6 is used for the whole guide in the simulation. For a ballistic design with $w_{max}=$\,10\,cm and a compression of 0.80, the formulae in \cite{NonLinearFocusingGuide} result in a non-linear guide part of $L_{NL}=$\,17.1\,m length followed by a $L_{lin}=$\,2.1\,m long linear guide. The guide width at the cross-over point is 4.67\,cm. Such a guide section is used as both the diverging as well as the converging part of the ballistic guide. Note that the non-linear focusing guide has only been calculated to retain a rectangular phase space without introducing inhomogeneities; its performance after a kinked or curved guide section, where the phase space cannot be expected to be rectangular, has not been studied before. 

The kinked version is designed using the same principles as in the \textit{extended kinked ballistic double-ellipse B} solution - with a central narrow point calculated from the kink angle chosen such that direct line of sight is avoided - which leads to a width of 4.1\,cm at the center of the guide.

\begin{figure}[tb!]
\centering
\subfigure[\label{f_DivNL}]{\includegraphics[width=.48\linewidth]{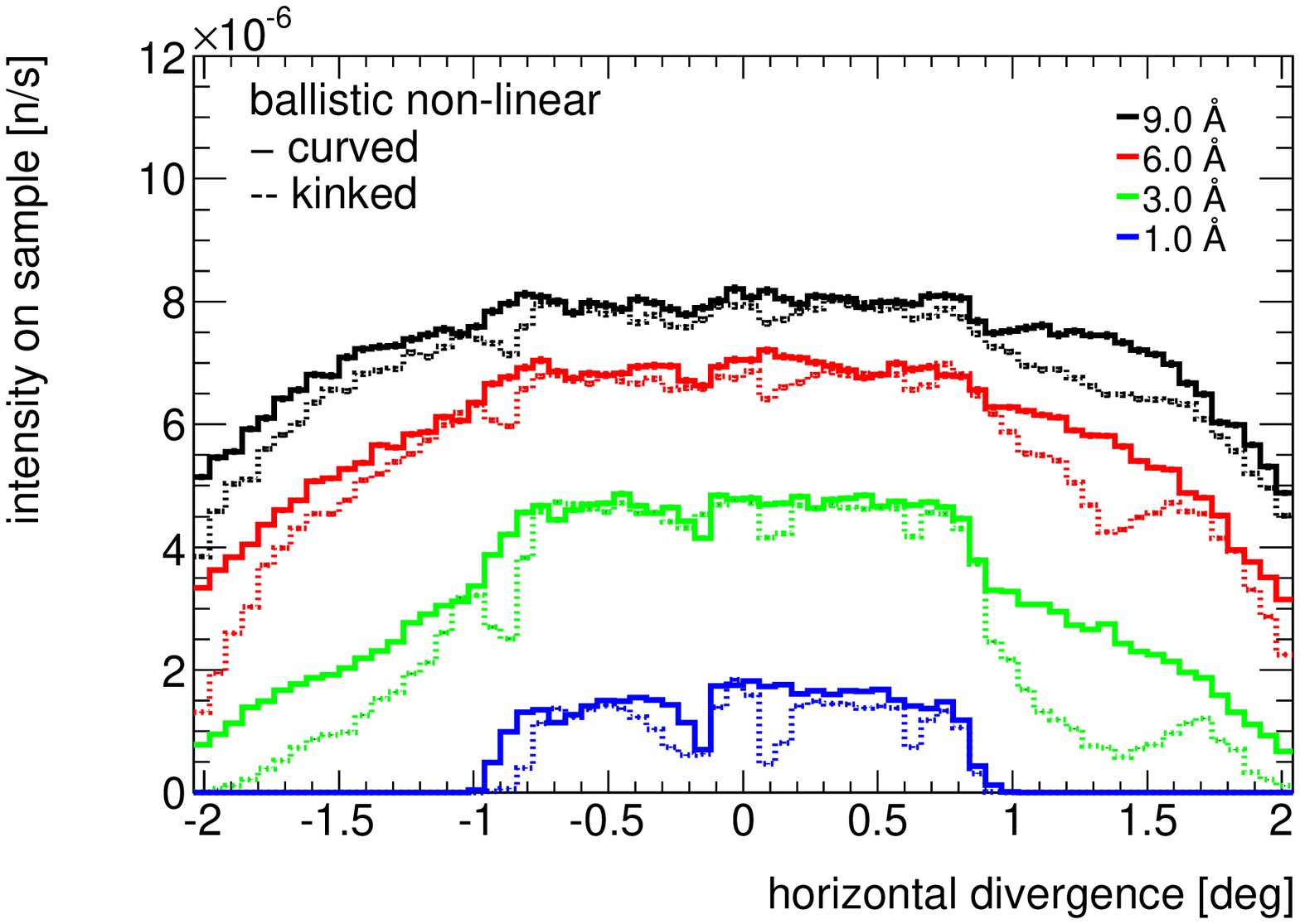}}
\subfigure[\label{f_BTratio_NL}]{\includegraphics[width=.48\linewidth]{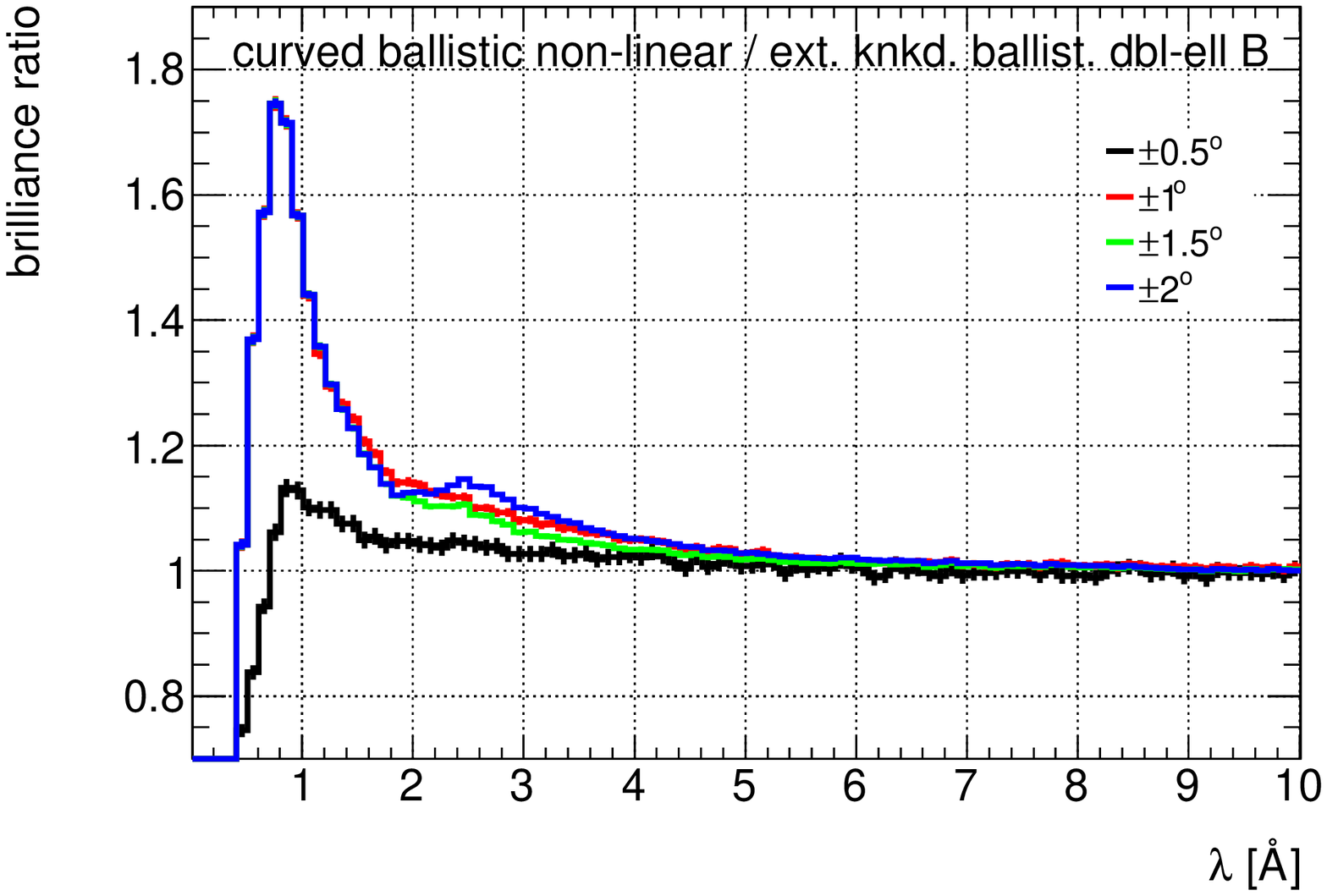}}
\caption{Performance of the non-linear guide design: horizontal divergence in bins of 0.06\DEG\,of curved (solid) and kinked (dotted) design (a), and brilliance ratio for different divergence regions compared to the \textit{extended kinked ballistic double-ellipse B} (b).}
\label{f_NL}
\end{figure}

Simulation shows that the curved option yields a smoother divergence and more intensity on the sample than the kinked version, see figure~\ref{f_DivNL}. The horizontal divergence of the curved option is comparable to the one obtained with the \textit{extended kinked ballistic double-ellipse B} solution for 6\,\Angst\,and 9\,\Angst, while shorter wavelengths show a slight asymmetry and there is more structure in the 1\,\Angst\,divergence spectrum. The brilliance on the sample is up to 70\,\% higher for very short wavelengths $\lambda \sim$0.8\,\Angst\,and divergences of 1\DEG\,or larger, see figure~\ref{f_BTratio_NL}. However, the difference in transmission decreases quickly with wavelength to less than 20\,\% for $\lambda>$\,1.5\,\Angst, and for longer wavelengths for which the beam homogeneity is comparable to the one obtained with the solution found here, the transmission is also not significantly different. 

\begin{figure}[tb!]
\centering
\subfigure[\label{f_PhaseSpace_G7pp}]{\includegraphics[width=.48\linewidth]{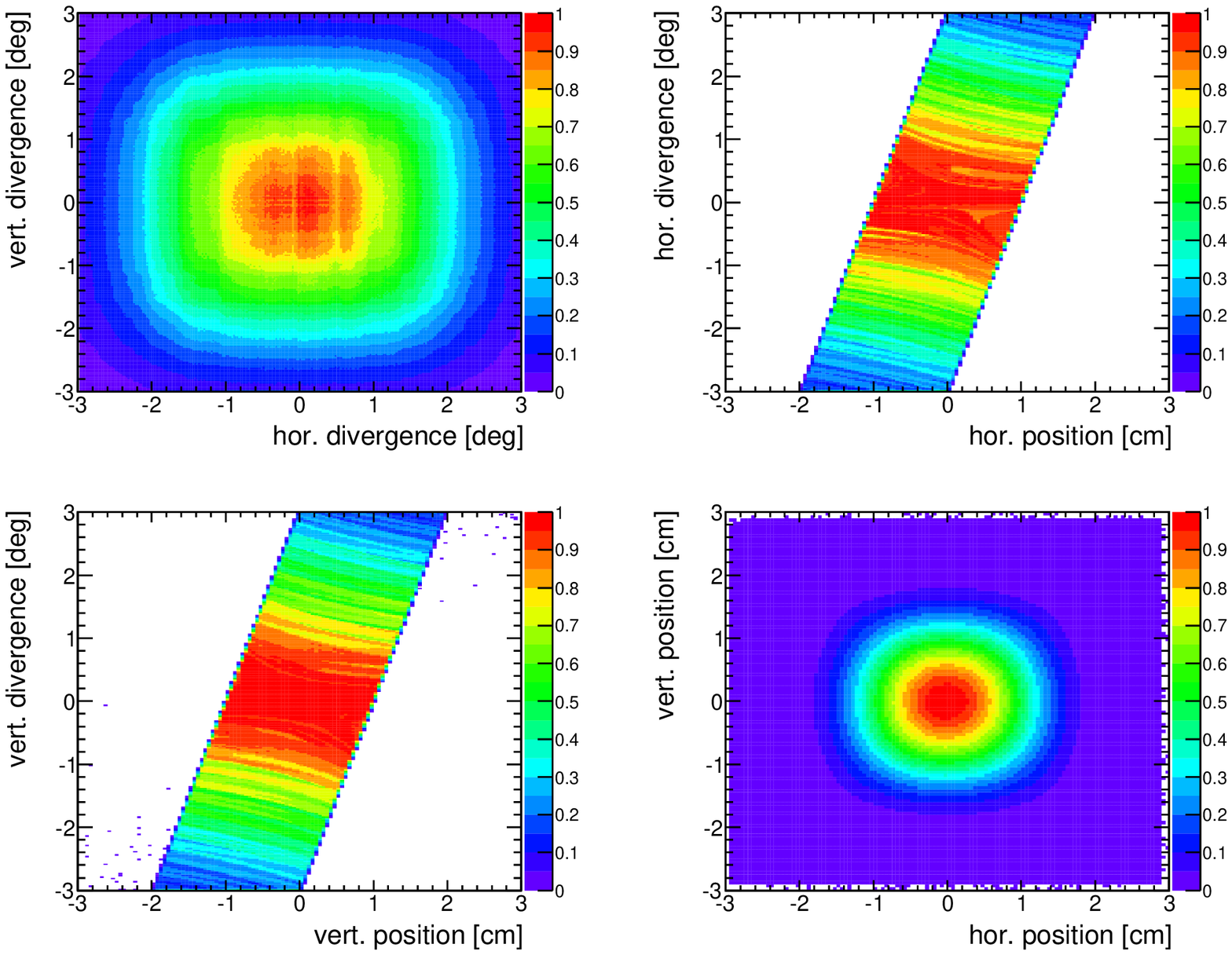}}
\subfigure[\label{f_PhaseSpace_NL}]{\includegraphics[width=.48\linewidth]{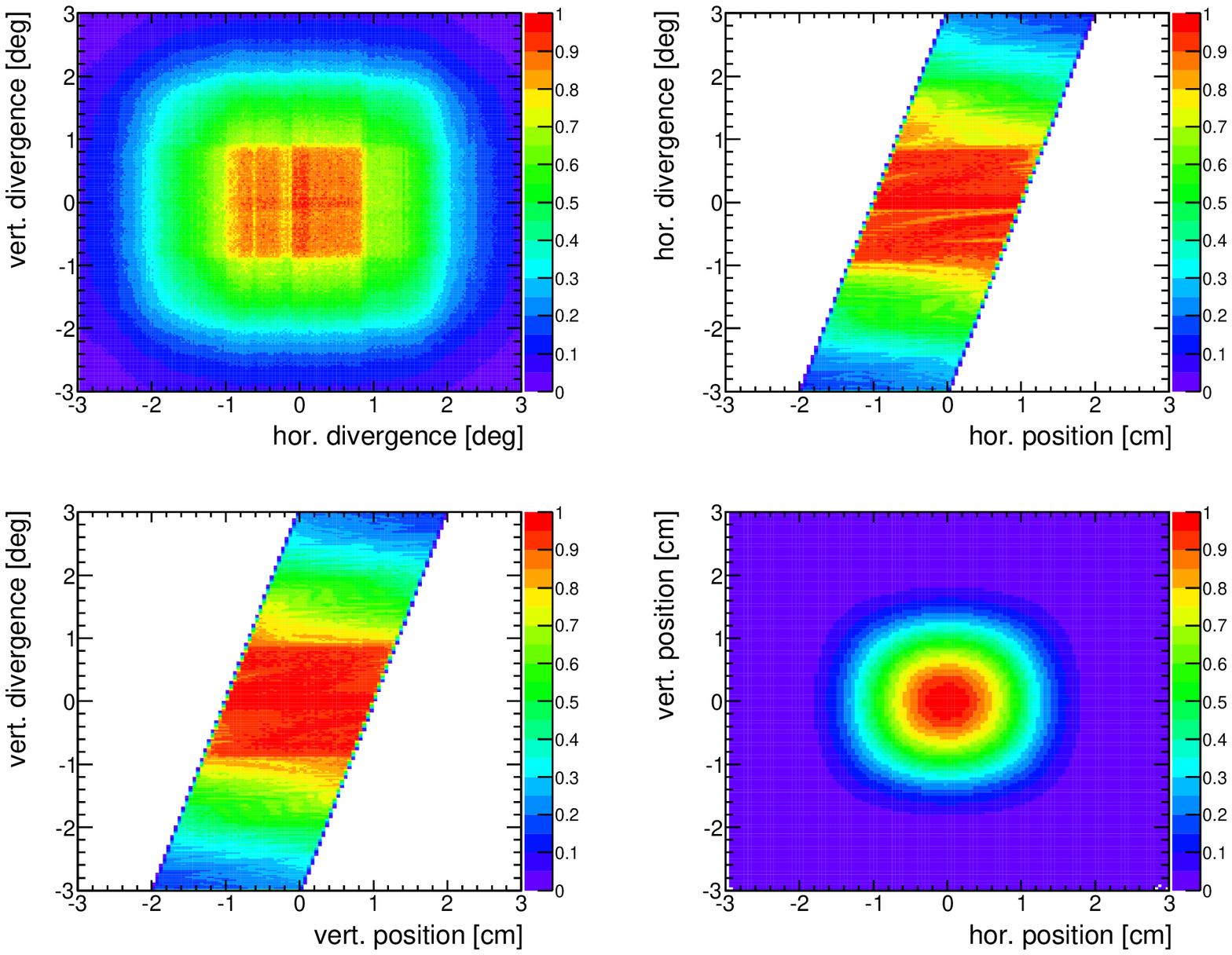}}
\caption{Two-dimensional divergence (upper left) and position (lower right) as well as horizontal (upper right) and vertical (lower left) phase space of the whole wavelength spectrum at the sample position, obtained with the \textit{extended kinked ballistic double-ellipse B} (a) or the curved non-linear guide (b). }
\label{f_PhaseSpace}
\end{figure}

For a better comparison of these two solutions, the two-dimensional phase space diagrams are shown in figure~\ref{f_PhaseSpace_G7pp} for the \textit{extended kinked ballistic double-ellipse B} and in figure~\ref{f_PhaseSpace_NL} for the analytical non-linear solution. The spatial focusing is very similar, while in the three other diagrams, the analytical solutions shows a slightly more rectangular distribution especially in the two-dimensional divergence. 

In summary, for long wavelength neutrons the performance of the curved non-linear guide is similar to the one of the \textit{extended kinked ballistic double-ellipse B}, while short wavelength neutrons can be delivered in higher abundance but with a more inhomogeneous divergence. For a (virtual) source not too small to illuminate the whole guide entry as used so far, the preferred guide shape thus depends on the relative importance of intensity versus beam homogeneity (the performance with a smaller source sizes is described in the next section). \newline

For both the hybrid as well as the non-linear alternative, it should be noted that only straight-forward guide designs, as similar to the solutions found here as possible, have been simulated. A numerical optimization might give parameters for slightly better results for all guide concepts studied here, but the overall tendency cannot be expected to change, and accounting for a homogeneous phase space in an optimization is challenging and will be treated elsewhere.

\section{Source size} \label{s_SourceSize}

\begin{figure}[tb!]
\centering
\subfigure[\label{f_DivSourceG7pp}]{\includegraphics[width=.45\linewidth]{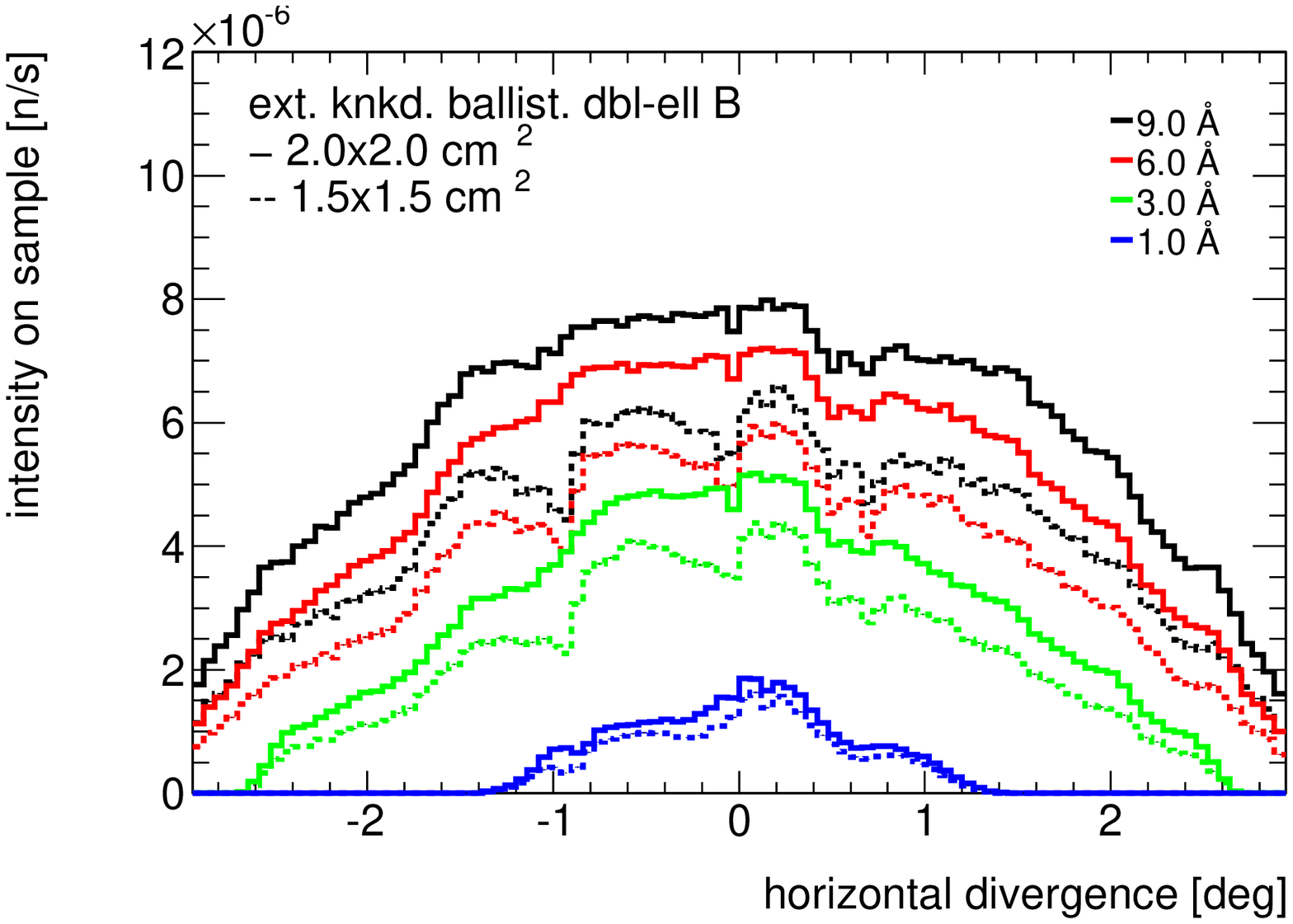}}
\subfigure[\label{f_DivSourceNL}]{\includegraphics[width=.45\linewidth]{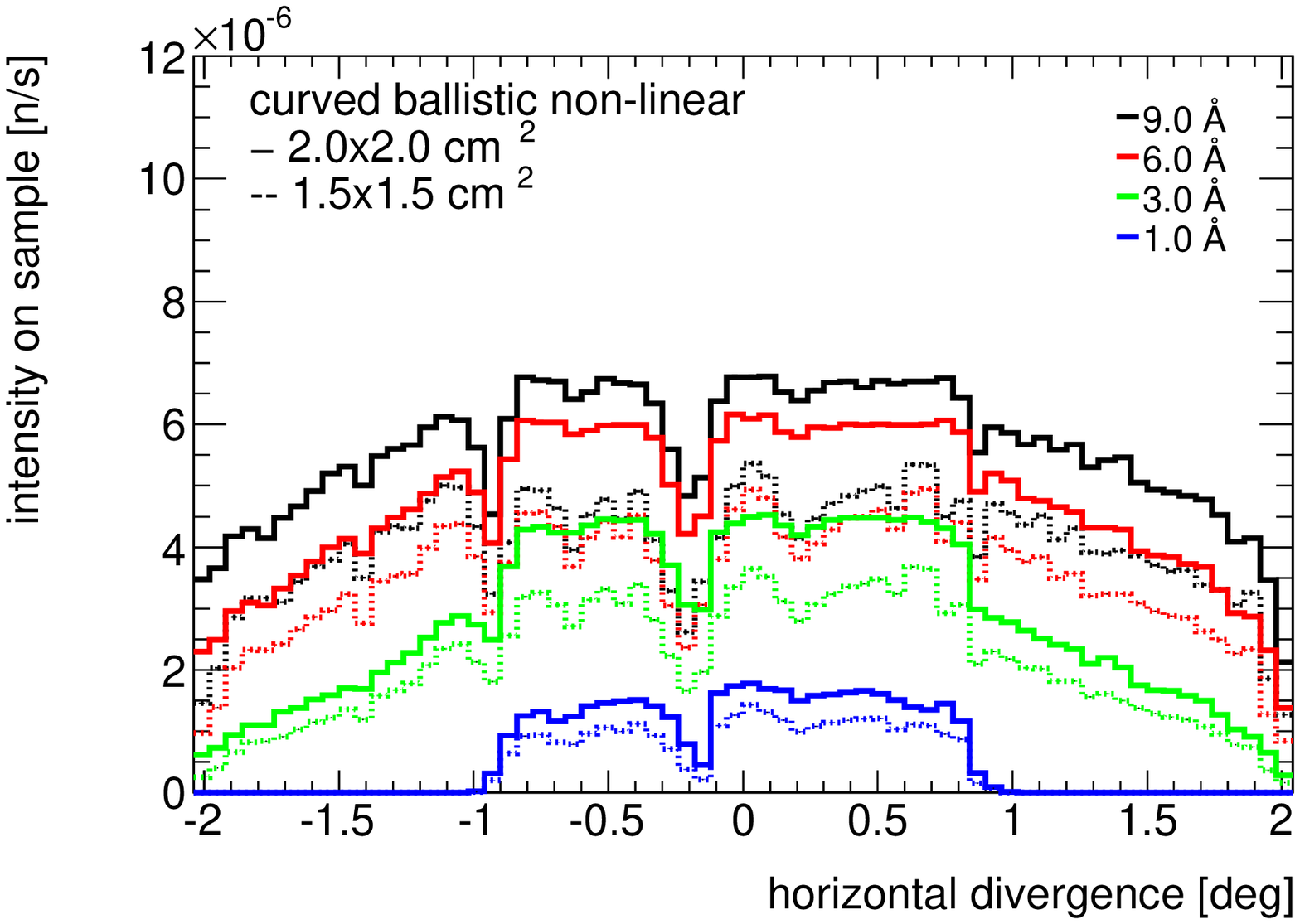}}
\caption{Horizontal divergence distribution in bins of 0.06\DEG\,with smaller source size for the \textit{extended kinked ballistic double-ellipse B} (a) and the curved non-linear (b) solutions.}
\label{f_DivSource}
\end{figure}

The phase space produced with an elliptically focusing guide at the sample position was shown to become more homogeneous when the guide entrance is decreased \cite{Schanzer200463}, and one recently found requirement for a smooth divergence profile with elliptic neutron guides is indeed a large source compared to the guide entry \cite{MultipleReflectionsInEllipticNeutronGuides}. A detailed study of gravity effects in elliptic neutron guides found a necessary ratio of source to guide entry width of about 1, with the exact value depending on the neutron wavelength \cite{GravityInEllipticGuides}. In the \textit{extended kinked ballistic double-ellipse B}, the same effect is observed: the solid lines in figure~\ref{f_DivSourceG7pp} show some minor structure in the divergence for a 2$\times$2\,cm\sq\,source the same size as the guide entry, while a pronounced structure is seen in the dotted lines which correspond to a 1.5$\times$1.5\,cm\sq\,source. 

The curved non-linear guide is more susceptible to smaller source sizes: figure~\ref{f_DivSourceNL} shows a significant structure already for a 2.0\,cm wide source, and with a 1.5$\times$1.5\,cm\sq\,source an inhomogeneity much larger than with the \textit{extended kinked ballistic double-ellipse B}, accompanied by a larger transmission loss for long wavelength neutrons. Therefore this solution is only compatible for medium sized sources; decreasing the guide entry for smaller sources will either lead to a larger compression factor and hence more structure in the divergence, or in combination with a decreased maximum width to a smaller transmission.

\section{Discussion}

A 150\,m long guide was developed out of a double-elliptic design by replacing the central part with a linear segment while keeping a central narrow point, first motivated by the possibility of a chopper placement but later seen to improve the divergence profile if used in the right way to reduce the kink angle necessary to avoid a direct line of sight from source to sample. With a maximal guide width of 15\,cm, simulation revealed the ideal transition point between elliptical and linear guide shape to be at a guide width of 8\,cm beyond the maximum of the ellipse, corresponding to 92\,\% of the guide length being elliptic. This \textit{extended kinked ballistic double-ellipse B} shows very good performance in terms of both brilliance transfer and beam homogeneity. 
 
While the proposed guide is rather simple using standard components, it was shown that recently proposed more complex guide designs don't give a significantly better performance: A ballistic hybrid guide yields lower beam homogeneity and slightly lower transmission of neutrons with a divergence smaller than $\pm$1\DEG. An analytical non-linear focusing guide put in a ballistic context with curved central section shows a similar performance for long wavelength neutrons and an increased transmission accompanied by decreased beam homogeneity for shorter wavelength. If the source size is the same value as the guide entry width or smaller, the divergence profile obtained with the curved analytical non-linear guide shows much larger inhomogeneity as the \textit{extended kinked ballistic double-ellipse} proposed here, and also the transmission decreases more with decreasing source size. Therefore for medium sized sources larger than the guide entry this option is compatible and even gives a larger intensity of very short wavelength neutrons, while it is less useful for smaller (virtual) sources. 

A large distance between guide exit and a small sample should however not be desired, since the analytical focusing guide does not have a focal point that can be placed in or behind the sample position, but yields a compressed beam directly at the guide exit. In the example used here, the wavelength integrated intensity on a 1$\times$1\,cm\sq\,sample at the sample position 19\,cm behind the guide is still 90\% of the intensity at the guide exit. In a distance of 50\,cm from the guide exit, however, the intensity has dropped to 30\%.

\section{Conclusions}

The 150\,m long extended kinked ballistic double-ellipse developed here gives a high brilliance transfer and a homogeneous divergence profile on the sample while avoiding direct line of sight between source and sample. If the (virtual) source size is not too small and the sample can be placed close to the guide exit, a curved ballistic guide with analytical non-linear focusing sections is an alternative to consider, in particular if high intensity of short wavelength neutrons is more important than a perfectly homogeneous beam divergence in that part of the wavelength spectrum.

\section{Acknowledgments}
We thank L. D. Cussen for sharing unpublished ideas and involving in extensive discussions.

This work was funded by the German BMBF under ``Mitwirkung der Zentren der Helmholtz Gemeinschaft und der Technischen Universit{\"a}t M{\"u}nchen an der Design-Update Phase der ESS, F{\"o}rderkennzeichen 05E10CB1''.

\label{Bibliography}
\bibliographystyle{unsrtnat}
\bibliography{References}
\begin{appendix}
\section{On the calculation of line of sight avoidance}

Whenever a kink angle or curvature radius necessary to avoid direct line of sight to the source is calculated, the guide width is increased by 2$t_g$ in the calculation to account for glass substrates on both sides of the guide of $t_g$ thickness each, which are known to be transparent for high energy particles. The glass substrate is assumed to be $t_g=$1\,cm in all calculations.

\subsection{Kinked ballistic guides}
\begin{figure}[tb!]
\centering
\subfigure[Guide wall on central focal point \label{f_LoS_kink}]{\includegraphics[width=.45\linewidth]{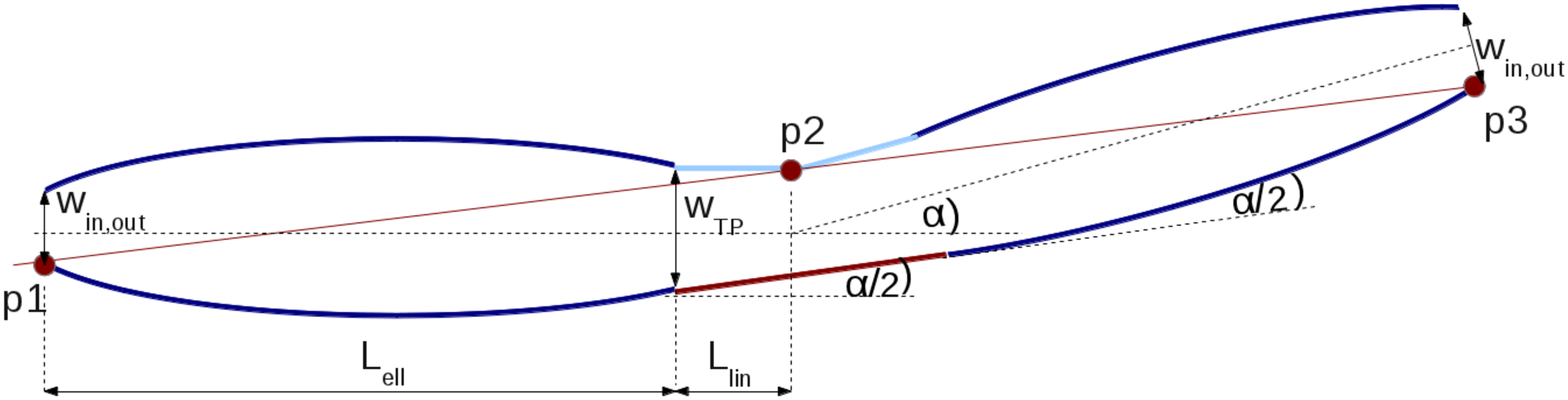}}
\subfigure[Guide wall inclination from line-of-sight \label{f_LoS_R}]{\includegraphics[width=.45\linewidth]{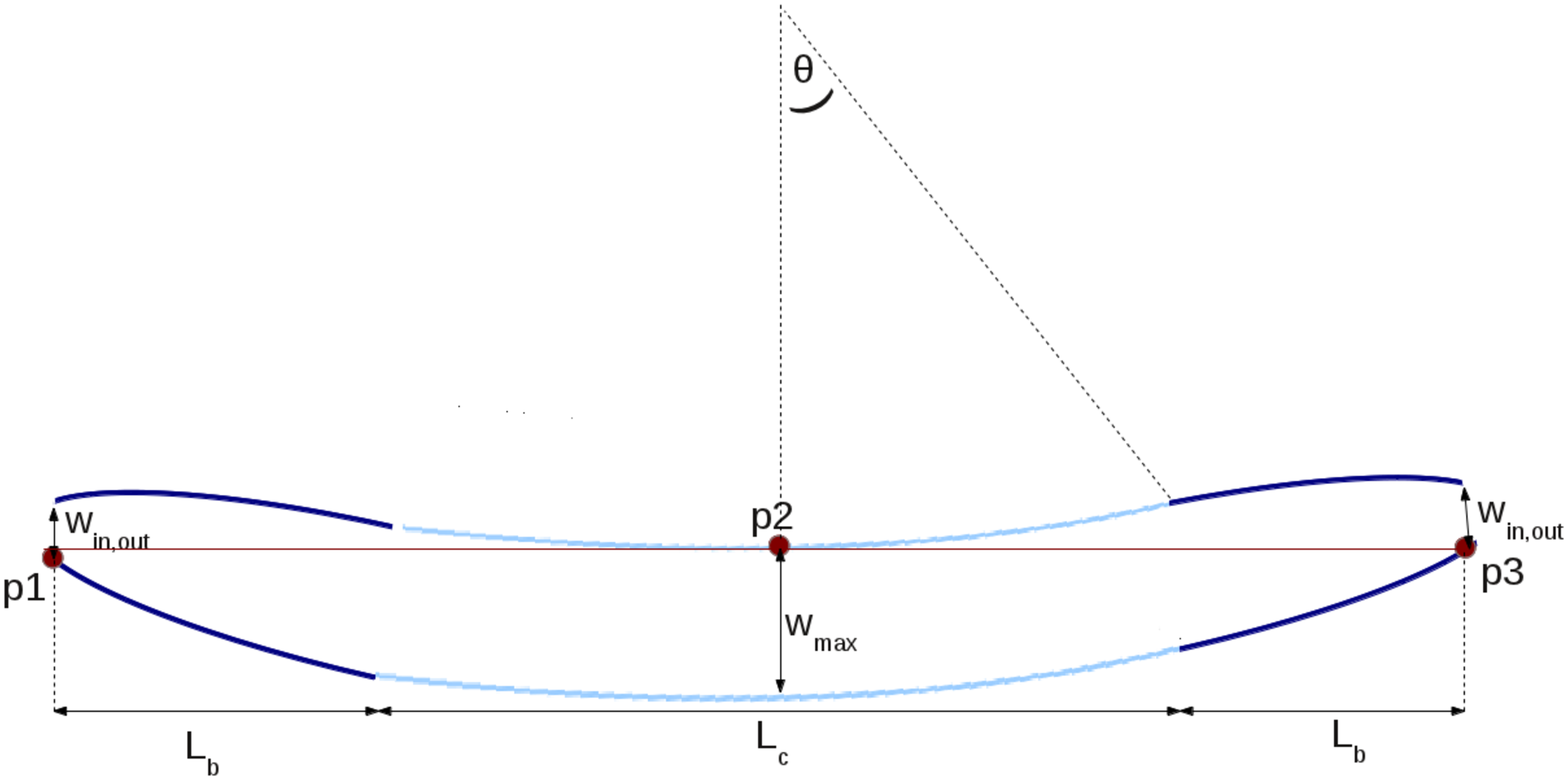}}
\caption{Schematic drawing of kinked and curved guide designs and neutron trajectory (red line) relevant for direct line of sight.}
\label{f_LoS}
\end{figure}

The kink angle necessary to avoid direct line of sight is determined by the three points marked red in figure~\ref{f_LoS_kink}: The possible neutron trajectory shown by the red line has an incoming divergence of $$\alpha_{in} = \arctan{\left(\frac{w_{in,out}/2+w_{TP}/2}{L_{ell}+L_{lin}}\right)}$$ (note that guide walls drawn in light blue are parallel to the guide axes), where $w_{TP}$ is the guide width at the transition point between elliptic and linear guide section. Hence its horizontal offset at the guide exit is $L_{tot} \tan{\left(\alpha_{in}\right)}$ with the total guide length $L_{tot}=2(L_{ell}+L_{lin})$. If a transparent glass substrate of thickness $t_g$ is assumed on each side of the guide, the angle changes to $$\alpha_{in} = \arctan{\left(\frac{w_{in,out}/2+w_{TP}/2+2t_g}{L_{ell}+L_{lin}}\right)} \mbox{    .}$$ 
The offset of the guide $(L_{ell}+L_{lin}) \tan{\left(\alpha\right)}$ has to exceed that, so with $\tan{\left(\alpha\right)}\simeq\alpha$, the kink angle $\alpha$ follows as $$\alpha/2 > \left( \frac{w_{in,out}+w_{TP}+4t_g}{L_{tot}} \right) \mbox{    .}$$

\subsection{Curved ballistic guides}
Similar to the kinked guide, the neutron trajectory through the points $p1$ and $p2$ in figure~\ref{f_LoS_R} determines the curvature needed to reach point $p3$. Due to the symmetry of this design, these points are the same as for a guide with constant cross-section of effective width $w_{eff}=w_{in,out}/2+w_{max}/2+2t_g$, where the last term again accounts for the glass substrate. The horizontal offset at the end of the curved section with respect to p2 is $\frac{1}{2}R\theta^2 =: \Delta y_c$, the additional offset by the following elliptic part is $L_b \tan{\left(\theta\right)} =L_b \frac{\Delta y_c}{L_c/4}$. Hence the horizontal offset between points 2 and 3 is 
$$ \frac{R}{2}\theta^2 \left(1 + \frac{L_b}{L_c/4} \right) = w_{eff} \mbox{    ,}$$ and with $\theta=\frac{L_c/2}{R}$ follows
$$ R=\frac{L_c^2}{8w_{eff}} \left( 1 + \frac{4L_b}{L_c} \right) \mbox{    .}$$

\end{appendix}

\end{document}